\documentstyle[12pt,epsfig]{article}
\def\gappeq{\mathrel{\rlap {\raise.5ex\hbox{$>$}}
{\lower.5ex\hbox{$\sim$}}}}
\def\permil{$\%\raise.20ex\hbox{$_0$}}
\def\lappeq{\mathrel{\rlap{\raise.5ex\hbox{$<$}}
{\lower.5ex\hbox{$\sim$}}}}

\begin{document}
\topmargin -1.0cm
\oddsidemargin -0.8cm
\evensidemargin -0.8cm
\pagestyle{empty}
\begin{flushright}
CERN-TH/97-203
\end{flushright}
\vspace*{10mm}
\begin{center}
{\Large\bf      Anomalous U(1),}\\
\vspace{0.5cm}
{\Large\bf     Gauge-Mediated Supersymmetry Breaking and}\\
\vspace{0.5cm}
{\Large\bf        Higgs as  Pseudo-Goldstone Bosons}\\
\vspace{2cm}
{\large\bf Gia Dvali$^{\, a}$  and Alex Pomarol$^{\, b}$}\\
\vspace{.7cm}
$^{a}${Theory Division, CERN}\\
{CH-1211 Geneva 23, Switzerland}\\
\vspace{0.3cm}
$^{b}${Institut de F\'\i sica d'Altes Energies}\\
{Universitat Aut\`onoma de Barcelona}\\
{E-08193 Bellaterra, Barcelona, Spain}\\
\end{center}
\vspace{2cm}
\begin{abstract}

We study the breaking of supersymmetry in models with
anomalous U(1). 
These models are simple to construct 
and contain 
natural candidates for
being the messengers of
gauge-mediated  supersymmetry breaking. 
When
some of the ordinary matter fields
transform under the anomalous U(1),
we find a 
  hybrid scenario in which  
the U(1) and the gauge interactions
mediate the breaking of supersymmetry. 
This leads to a hierarchy of soft
masses between the charged and neutral fields and provides 
a solution to the  $\mu$-problem.
Among these models, we present a scenario in which
 the Higgs arises as a pseudo-Goldstone 
boson. 
This scenario naturally allows for
values of the  $\mu$-term and the scalar soft masses 
larger  than the weak scale.

\end{abstract}

\vfill\vspace{1cm}
\begin{flushleft}
CERN-TH/97-203\\
August 1997
\end{flushleft}
\vfill
\eject
\pagestyle{empty}
\setcounter{page}{1}
\setcounter{footnote}{0}
\pagestyle{plain}
 

\section{Introduction}

Anomalous U(1)   
allow for simple models of supersymmetry 
breaking \cite{us,bd}. 
The presence of a gravitational anomaly for the U(1), 
${\rm Tr}{\bf Q} \not= 0$  (${\bf Q}$  being the 
U(1)-generator), 
 is  necessary  for the generation of 
a   Fayet-Iliopoulos (FI) term and  the triggering
of  supersymmetry breaking.
In effective field theories arising
from string,   an anomalous U(1) is often present. 
Interestingly enough, one finds that   these theories 
always  contain  fields that transform simultaneously under 
the anomalous U(1) and the standard model (SM) group.
 This is a necessary condition 
to cancel the anomaly by the Green-Schwarz mechanism \cite{gs}.

This allows for two possibilities 
to transmit the breaking of supersymmetry
to the SM fields:

 ({\it i}) The  quiral superfields of the SM
(quarks, leptons and Higgs)
are neutral under the
anomalous U(1). 
In this case, there should be extra matter
 fields (to cancel the mixed anomalies)
that transform
nontrivially under the U(1).
These fields can act as messengers
and communicate the breaking of supersymmetry 
to the SM fields at   a higher loop level. 
(In ref.~\cite{us} this effect was the
main source for the gaugino masses).
This will  correspond 
to ordinary models of gauge-mediated supersymmetry
breaking \cite{gmsb} \footnote{In a different context the possibility of
 gauge-mediation with anomalous U(1) was considered in \cite{zura}.}.

  ({\it ii}) There are SM fields  transforming 
under the anomalous U(1).
These models will present  a hierarchy of  the soft masses;
fields charged under the U(1) will get tree-level  scalar masses, while
neutral ones and gauginos will get soft masses at higher loop orders.
We will see that the tree-level masses are renormalization-scale invariant
and are therefore (almost) not modified by radiative corrections.
Consequently, 
they are not affected by the physics at high energies.
Models of this type have already been 
considered in refs.~\cite{us,models,zura}.
In particular, in ref.~\cite{us}  we proposed a model with
 the first and second generations
of quarks and lepton charged under the anomalous U(1) as a solution
to the flavor and CP supersymmetric problem.

Here we will  present models of the type
({\it i}) and ({\it ii}), and compare the different
  pattern of soft masses that they generate.
We will
 consider models with the Higgs transforming under the 
anomalous U(1) and
 show that  this case allows for  
 a new solution to the $\mu$-problem.
Furthermore, these models can easily implement scenarios in which 
one of the  Higgs arises 
as a pseudo-Goldstone boson (PGB) associated with the
 breaking of  an accidental symmetry of the Higgs potential \cite{pgbmodels}.
We will study this scenario and show that it leads naturally to
values of $\mu$ and  soft
masses larger than the weak scale.
The origin of this
hierarchy ($\mu,m_{H,Q}> m_Z$) can be understood from symmetry principles
\cite{pgbbefore}.
The weak scale $m_Z$ 
is determined by the PGB mass that, due to the
accidental symmetry of the Higgs potential,
is smaller than the other mass parameters (soft supersymmetry breaking
masses) of the Higgs potential. 
This accidental symmetry is not preserved beyond the tree-level
approximation 
and then one does not expect  that the hierarchy will be
stable under radiative corrections. 
We find, however, that, 
due to a partial cancellation between the one-loop corrections,
  $\mu$ is maintained much  larger than $m_Z$.
In this scenario 
the scalar superpartners and Higgsinos will be heavier than expected
and not accessible  at LEPII.

In section~2 we will give the conditions for supersymmetry breaking
in theories with anomalous U(1). In section~3 we will show that 
the scalar soft masses are renormalization-scale invariant
and will study its consequences.
In section~4 we will present a  model with gauge-mediated
supersymmetry breaking and compare it with models in which the
breaking of supersymmetry is also communicated by the anomalous U(1).
In section~5  we show how the $\mu$-term can arise in these models.
We also present a model in which one of the Higgs is a  PGB and study how
large the $\mu$-term can be.
Section~6 is devoted to conclusions.
We include an appendix in which  we show an explicit model 
with dynamical supersymmetry breaking. 

\section{Supersymmetry breaking with  an anomalous U(1)}

Let us  consider a U(1) theory with  
${\rm Tr}{\bf Q} \not= 0$ 
 and  therefore that it is anomalous. 
At the one-loop level, this
 results \cite{witten} into the appearance of a tadpole $D$-term,  
the Fayet-Iliopoulos (FI) 
term, $\xi$. 
This tadpole is quadratically divergent and thus 
   $\xi\sim\frac{1}{16\pi^2}\Lambda^2{\rm Tr}{\bf Q}$
where $\Lambda$ is the 
cut-off of the theory.
In string theories 
  the generated FI-term 
can be calculated with a stringy regularization
and is given by \cite{fayetil}
\begin{equation}
\xi=\varepsilon M^2_P\, ,  
\end{equation}
where
\begin{equation}
 \varepsilon= {g^2{\rm Tr}{\bf Q} \over 192\pi^2}
\left(\frac{\sqrt{2}M_{st}}{gM_P}\right)^2\, ,
\label{varepsilon}
\end{equation}
where $M_{st}=1/\sqrt{\alpha^{\prime}}$ is the string scale. 
In the weakly-coupled heterotic string, $M_{st}$   is related to 
the reduced Planck scale $M_P\simeq 2\times 10^{18}$ GeV by 
$M_{st}=gM_P/\sqrt{2}$ and therefore we have $\varepsilon\sim 10^{-3}$
for ${\rm Tr}{\bf Q}={\cal O}(1)$.

Let us also consider that exists 
a pair of chiral superfields
$\phi$ and $\bar\phi$ with charges equal to $-1$ and $q_{\bar\phi}$
 respectively under the anomalous U(1).
The $D$-term contribution of the U(1)
to the effective potential takes the form
\begin{equation}
 {g^2 \over 2}D^2 = {g^2 \over 2}
\left(q_{\bar\phi}|\bar\phi|^2 - |\phi|^2+\xi\right)^2\, .
\label{dtermcon}
\end{equation}
If eq.~(\ref{dtermcon}) is the only term in the potential,
the vacuum expectation value (VEV) of $\phi$  adjusts
 to compensate $\xi$ (that we assume to be positive), 
and supersymmetry will not be broken.
Nevertheless, as pointed out by
 Fayet  long time ago \cite{fayet},
supersymmetry can be spontaneously broken 
if the
field $\phi$  has a nonzero mass term in the superpotential:
\begin{equation}
W = m\phi\bar\phi\, .
\label{super}
\end{equation}
Notice that $m$ carries U(1)-charge equal to $(1-q_{\bar\phi})$ that 
can be different from zero if $m$ arises, as we will see, 
from some nonperturbative
dynamics.
In this case $m$ should have a nontrivial dilaton dependence
\begin{equation}
m \sim {\rm e}^{-S\frac{192\pi^2}
{ {\rm Tr}{\bf Q}}(1-q_{\bar\phi})}M_P \, ,
\end{equation}
in the normalization in which 
the dilaton shifts under a U(1)-transformation
as $S \rightarrow S + i \alpha {{\rm Tr}{\bf Q} \over 192\pi^2}$ and
$\phi \rightarrow {\rm e}^{i\alpha}\phi$.

For the moment, let us consider   $m$ as a parameter whose value
is much
smaller than  $M_P$. 
As we will show, the mass term $m$ has to lay $\sim 1$ TeV to generate
realistic masses.
Minimizing
the potential derived from
 eqs.~(\ref{dtermcon}) and (\ref{super}),
we obtain that the VEVs of the scalar components are 
\begin{equation}
\langle\bar\phi\rangle = 0,~~~\langle\phi\rangle^2 
= \xi- {m^2 \over g^2}\, ,
\label{vacuums}
\end{equation}
and the VEVs of the $F$- and $D$-components are given by
\begin{equation}
\langle F_{\bar\phi}\rangle = m\sqrt{\xi- {m^2 \over g^2}},~~~
\langle F_{\phi}\rangle = 0,~~~ \langle D\rangle =\frac{m^2}{g^2}\, .
\label{vacuum}
\end{equation}
A important feature of these models is that 
supersymmetry is broken by nonzero VEVs of 
the two auxiliary fields, $F$ and $D$.  

Embedding the above model in a supergravity theory will not restore 
supersymmetry as it was proven in ref.~\cite{us};
the VEVs of
eqs.~(\ref{vacuums}) and (\ref{vacuum}) are just shifted
by the gravity corrections.
The only noticeable effect is 
a tadpole for $\bar\phi$  generated from the bilinear
soft-term $\approx m_{3/2}m\langle\phi\rangle\bar\phi$ 
where $m_{3/2}\simeq \langle F_{\bar\phi}
\rangle/M_P$ 
is the gravitino mass.
Thus, $\bar\phi$ gets a nonzero VEV given by
\begin{equation}
\langle\bar\phi\rangle\simeq\frac{ m_{3/2}}{m}\langle\phi\rangle\simeq
\sqrt{\varepsilon}\langle\phi\rangle\, .
  \label{newvev}
\end{equation}
The fact that the VEV of $\bar\phi$ is smaller than that of $\phi$
(since $\varepsilon\ll 1$)
 will have important phenomenological consequences.

Let us comment on the origin of the small mass-parameter  $m$. 
It can be  generated dynamically \cite{us,bd}.
For example, it can arises  from  a  
field  condensation, $m=\langle\Phi\bar \Phi\rangle/M_P$,
where 
$\Phi$ and $\bar\Phi$ are some fields that
 transform under a strongly interacting gauge group \cite{nonpert}.
In section~4 and the appendix we show some examples. 
If $m$ is dynamical, the minimization of the full potential 
is slightly different from
 the case above in which we took $m$ as a frozen parameter.
We find (see appendix) that supersymmetry is always broken,
 although the  VEV of
$\bar\phi$ cannot be exactly determined  since it depends on
unknown nonperturbative effects.
We, however, estimate that $\langle\bar\phi\rangle$ takes a value
 between
$\sqrt{\varepsilon}\langle\phi\rangle$ and 
$\langle\phi\rangle$.

Alternatively, one can consider that $m$ arises from a
higher dimensional operator suppressed by  powers of the Planck scale.
For example,  a superpotential term such as
\begin{equation}
W=\frac{\phi^{12}\bar\phi}{M^{10}_P}\, ,
  \label{higher}
\end{equation}
 leads to an effective $m$ of order
\begin{equation}
  \label{effmu}
m\simeq\frac{\langle\phi\rangle^{11}}{M_P^{10}}
\sim\varepsilon^5\sqrt{\xi}\sim 1\ {\rm TeV}\, ,
\end{equation}
for $\varepsilon\simeq 10^{-3}$. 
Notice that this option does not need for
nonperturbative dynamics 
to generate a scale smaller than $M_P$.
This is possible due to the fact that the
 FI-term is generated at the 
one-loop level and this introduces a new scale in
the theory $\sqrt{\xi}\ll M_P$.
Although possible, we do not find the latter approach 
very appealing since it requires 
a very peculiar charge assignment $q_{\bar\phi} = 12$.

\section{Implications for the scalar soft masses}

Any scalar  field in the theory
 charged under the anomalous U(1) receive soft supersymmetry
breaking masses  
from the VEV of the $D$-term:
\begin{equation}
 m^2_{Q_i}=q_i\, m^2\, ,
\label{dtermmass}
\end{equation}
where $q_i$ is the U(1)-charge of $Q_i$.
The pattern of soft masses of eq.~(\ref{dtermmass}) 
has very interesting consequences.
First, we  have scalar mass degeneracy for fields with equal charges.
Second,  fields with   trilinear couplings 
$hQ_1Q_2Q_3$ in the superpotential has $q_1+q_2+q_3=0$,
and then
their  soft masses 
 will fulfill the following sum rule:
\begin{equation}
m^2_{Q_1}+m^2_{Q_2}+m^2_{Q_3}=0\, .  
\label{sumrule}
\end{equation}
The renormalization group equations (RGE) 
of the soft masses is given by
\footnote{We are assuming that Tr$[{\bf QY}]=0$ where ${\bf Y}$
is the hypercharge generator \cite{us}.}
\begin{equation}
\frac{dm^2_{Q_i}}{dt}=\frac{h^2}{8\pi^2}  
(m^2_{Q_1}+m^2_{Q_2}+m^2_{Q_3}+A^2)
-\frac{2C^a_i\alpha_a}{\pi}m^2_{\lambda_a}
\, ,
\end{equation}
where $C^a_i$ is
the quadratic Casimir  of the $Q_i$ 
representation of the group $G_a$, $A$ is the trilinear soft term
and $m_{\lambda_a}$ are the gaugino masses.
Therefore, 
the sum rule (\ref{sumrule}) 
implies that the soft masses are RG invariant 
up to corrections of ${\cal O}(m^2_{\lambda_a},A^2)$. 
These corrections, however, are small since  gaugino masses and 
trilinears do not receive contributions 
from the $D$-terms; 
 they will be generated from gravity contributions as in ref.~\cite{us},
or 
at a  higher loop-order 
as we will see in the next section.

The above  has  important phenomenological implications. 
If  we 
 assume that  the first and second families are
equally charged under the U(1), then the squarks and sleptons 
will be degenerate at the scale $\sqrt{\xi}$.
The running from the scale $\sqrt{\xi}$ to lower scales will not
modify 
the scalar masses,
since, as we said, they are renormalization-scale invariant
for $m_Q\gg m_{\lambda},A$. 
Therefore,
even 
if at a high energy all
the Yukawa coupling are of ${\cal O}(1)$ as in certain 
theories of flavor \cite{hkr},
the  squarks will keep their degeneracy at lower energies. 
This property can avoid the supersymmetric flavor problem \cite{hkr}. 
 These models  do not suffer from
the effect of ref.~\cite{hkr},
even though the supersymmetry breaking 
is communicated to the SM fields
at high energies.

Of course, this will be true up to gravitational corrections.
The squark masses receive 
gravity contributions that,
for nonminimal  K\"ahler potential, are nonuniversal \cite{soni}. 
These contributions 
are of the order of the gravitino mass
\begin{equation}
  \label{gravitino}
m^2_{3/2}\simeq \frac{\langle F_{\bar\phi}\rangle^2}
{M^2_P}\simeq \varepsilon m^2 \, .
\end{equation}
Therefore, gravity effects to eq.~(\ref{dtermmass}) are smaller than $1\%$ for 
$\varepsilon\lappeq 10^{-2}$.

\section{Gauge-mediated supersymmetry breaking}

Let us consider that the ordinary
SM fields
are neutral under the anomalous U(1). 
In this case, to 
cancel the anomalies by the Green-Schwarz mechanism \cite{gs},
the theory must contain
extra matter fields transforming under the
anomalous U(1). 
Let us consider that these extra
matter fields, $\Psi$ and $\bar \Psi$, are vector-like under the SM
group but  chiral 
under the anomalous U(1) ($q_\Psi+q_{\bar \Psi}\not=0$).
If  $n=-(q_\Psi+q_{\bar \Psi})/q_{\bar\phi}$ is positive, then 
 $\Psi\bar \Psi$
will couple to $\bar\phi$ \footnote{Equivalently, if
$q_\Psi+q_{\bar \Psi}$ is positive, then 
$\Psi\bar \Psi$ will couple to  $\phi$. The effects of this coupling
will be similar to those discussed below. }:
\begin{equation}
W=\frac{\bar\phi^n}{M_P^{n-1}}\Psi\bar \Psi\, .
\label{messengers}
\end{equation}
This  coupling will not change the vacuum 
of eqs.~(\ref{vacuums}) and (\ref{vacuum});
the supersymmetry breaking vacuum, 
however, will  be now a  local minimum. 
The scalar-component VEV of  $\bar\phi$ will give 
a supersymmetric mass to $\Psi$ and $\bar \Psi$,
while  the 
$F$-component VEV of  $\bar\phi$ will 
 induce a  mass splitting 
inside the scalar components of $\Psi$ and $\bar \Psi$.
Therefore, $\Psi$ and $\bar \Psi$ can act
as messengers and transmit 
the supersymmetry breaking to the SM fields.
Like ordinary  models with gauge-mediated supersymmetry breaking,
the gaugino masses will be given at the one-loop level \cite{gmsb}
\begin{equation}
m_{\lambda_a}=\frac{\alpha_a}{4\pi}nS^a_\Psi
\frac{\langle F_{\bar\phi}\rangle}{\langle\bar\phi\rangle}\, ,
\label{gauginos}
\end{equation}
where $S^a_\Psi$ is the Dynkin index of the $\Psi$-representation
of the gauge group $G_a$.
Scalar masses will arise 
 at the two-loop level 
\begin{equation}
{ m}^2_{Q_i}=2C^a_in^2S_\Phi\left( \frac{\alpha_a}{4\pi}\right)^2
\frac{\langle F_{\bar\phi}\rangle^2}{\langle\bar\phi\rangle^2}\, .
\label{squark}
\end{equation}
This contribution  is
universal for fields with equal quantum numbers;
therefore it does not induce flavor-violating interactions.
Trilinear and bilinear soft terms are  generated at the two-loop 
level, $A\simeq\frac{\alpha}{4\pi}m_\lambda$, unless there is
a messenger-matter mixing that will induce them at 
the one-loop level \cite{tri}.
Eqs.~(\ref{gauginos}) and (\ref{squark}) depend on the ratio
\begin{equation}
\frac{\langle F_{\bar\phi}\rangle}{\langle\bar\phi\rangle}=
\frac{\langle {\phi}\rangle}{\langle\bar\phi\rangle}m\, .
\end{equation}
We will therefore consider two separate scenarios:

\noindent\underline{Scenario (a)}: 
The VEV of $\bar\phi$ and $\phi$ are of the same order
and we have
\begin{equation}
  \label{relationa}
\frac{\langle F_{\bar\phi}\rangle}
{\langle\bar\phi\rangle}
\simeq m\, .
\label{scenarioa}
\end{equation}

\noindent\underline{Scenario (b)}: The VEV of $\bar\phi$ 
is smaller than the VEV of $\phi$ as in 
eq.~(\ref{newvev}).
We then have
\begin{equation}
  \label{relationb}
\frac{\langle F_{\bar\phi}\rangle}
{\langle\bar\phi\rangle}
\simeq\sqrt{\frac{1}{\varepsilon}}m\, .
\label{scenariob}
\end{equation}
Scenario (b) arises 
when 
$m$ is considered a  spurion field with a frozen  VEV
or when $m$ is effectively generated  from a
higher dimensional  operator such
as eq.~(\ref{higher}). 
When $m$ is induced dynamically 
by a field condensation (as in the appendix), 
we can have both scenarios
(see section~2).

Scenario (a) and (b) lead to different pattern of soft masses, as 
it is shown in
table~1.
To have $m_{Q}\sim 100$ GeV, 
 we need $m\sim 100$ TeV and $\sim 10$ TeV respectively
  for the scenario (a) and (b). 
There are also, however, 
gravity contributions to the scalar masses that 
can be comparable to the gauge contributions depending on the scenario
--see table~\ref{table}.
{}From eqs.~(\ref{gravitino}) and (\ref{squark}), 
we  have using eqs.~(\ref{scenarioa}) and (\ref{scenariob}) respectively
\begin{equation}
\frac{m^2_{3/2}}{m^2_Q}\simeq
\left\{\begin{array}{ll}
{10^{4}\varepsilon}&\ {\rm scenario\ (a)}\, ,\\
{10^{4}\varepsilon^2}&\ {\rm scenario\ (b)}\, .\\
\end{array}\right.
\label{nondeg}
\end{equation}
To avoid flavor problems coming from
gravity contributions, the ratio of eq.~(\ref{nondeg})
has to be smaller than $\sim 10^{-2}$.
We see that only for the
 the scenario (b) this can be satisfied
for values of $\varepsilon\sim 10^{-3}$.
Of course, for string theories with
$M_{st}\ll gM_P$  we see from eq.~(\ref{varepsilon})
that  both scenarios can accommodate
$\frac{m^2_{3/2}}{m^2_Q}\lappeq 10^{-2}$.

\subsection{A Complete model of dynamical supersymmetry breaking}

Here we present a  model of gauge-mediation that, at the same time,  
addresses the problem of the generation of the small scale $m$.
The model has the minimal gauge group and field content necessary to break
supersymmetry dynamically and transmit it by gauge interactions to the SM
fields \cite{otherat,ddrg}.

Take the group SU(N)$\times$U(1)$\times$SU(5)$_{SM}$, where the 
SU(N) becomes
strong at some scale $\Lambda\ll M_P$, the U(1) is anomalous
  and the SU(5)$_{SM}$ includes the SM group.
Add to the SM fields, the following field content:
\begin{eqnarray}
\phi&\ \  \ ({\bf 1},-1,{\bf 1})\, ,\\
\Psi&\ \ \  ({\bf N}, {1 \over 2},{\bf 5})\, ,\\
\bar\Psi&\ \ \  ({\bf \bar N},{1 \over 2},{\bf \bar 5})\, ,
\label{fields}
\end{eqnarray}
with the  classical superpotential
\begin{equation}
W=\lambda\phi\bar\Psi\Psi\ .
\end{equation}
This simple model leads to dynamical supersymmetry breaking 
\footnote{As far as supersymmetry breaking is concerned, this example
 is similar to that of ref.~\cite{bd}.}
with 
$\Psi$ and $\bar\Psi$ playing the role of condensates and messengers.
To show that,
let us 
ignore, for the moment, the $D$-term of the anomalous U(1).
Classically the
vacuum manifold (moduli space) has a flat direction parameterized by
the expectation value of $\phi$. Along this branch  $\Psi$ and  
$\bar{\Psi}$ are  massive,  $M_{\Psi}=\lambda \phi$,
 and can be integrated out. 
For $\phi \geq \Lambda$ the low-energy
theory is a pure super-Yang-Mills theory
with  a massless chiral superfield $\phi$ (and, of
course, all the SM states). 
Gaugino condensation
in the
strongly coupled group, $\langle\lambda\lambda\rangle$,
induces the nonperturbative superpotential
\cite{tvy,nsvz}
\begin{equation}
 W = \langle{\lambda}\lambda\rangle = N\Lambda^3_L\, ,
\end{equation}
where $\Lambda_L$ is the low-energy scale of the SU(N) theory. 
This scale is $\phi$-dependent; 
the explicit dependence can be found by the 
one-loop matching of the gauge coupling 
in the low and the high energy theories at the scale $\phi$ \cite{sv}.
By doing
this we arrive to the following effective low-energy superpotential:
\begin{equation}
W_{eff}=N \left (\lambda^5\phi^5\Lambda^{3N - 5}\right )^{\frac{1}{N}}\, .
\label{efsuperw}
\end{equation}
The superpotential (\ref{efsuperw}) leads to
spontaneous supersymmetry breaking  if
 $\phi$ gets a nonzero VEV. 
This is the case as soon as 
the anomalous $D$-term is switched-on
\footnote{In effective field theories arising from strings
the anomalous U(1) 
is also broken by the dilaton
VEV and the eaten-up component is a combination of $\phi$
and the dilaton; the
remaining combination is a light field. 
So strictly speaking, the question becomes
linked to the
story of dilaton stabilization, 
which we will not analyze here;  we are assuming
 that the dilaton is stabilized by some dynamics which 
{\it per se} preserves  supersymmetry.}.
The pattern of the induced soft masses  is like that in the scenario (a).

The superpotential (\ref{efsuperw})
 can also be derived 
by adding the instanton-generated 
 superpotential (for $N > 5$) \cite{ads}
\begin{equation}
W =(N-5) \left ({\Lambda^{3N - 5} 
\over {\rm det}M} \right )^{\frac{1}{N - 5}}\, ,
\end{equation}
and solving the equations of motion for the mesons $M^j_i =
 \bar\Psi^j {\Psi}_i$ (see appendix).
For $N = 5$ the same can be done but using
the quantum modified constraint \cite{nonpert}
\begin{equation}
W = {\cal A} \left( {\rm Det}M - B\bar B - \Lambda^{10} \right)\, ,
\end{equation}
where $B = {\rm Det} \Psi$ and
 $\bar B = {\rm  Det} \bar{\Psi}$ are baryons.
Note that if  all the other states   charged under the SU(5)$_{SM}$
are neutral under the U(1), 
then  anomaly cancellation 
by the Green-Schwarz mechanism \cite{gs}
requires  $N = 5$.
This case ($N = 5$)  shares some similarity 
with the models of ref.~\cite{ddrg}. 
The only difference resides  in the mechanism that
stabilizes $\phi$ at some large value. 
In the models of ref.~\cite{ddrg} 
this stabilization was achieved by 
the K\"ahler renormalization as in the 
Witten ``inverse hierarchy'' model \cite{wittenb}. 
Due to the logarithmic nature
of this effects the stabilization is only possible for $N = 5$ (when
the superpotential is linear). 
In our case $\langle\phi\rangle \gg \Lambda$ is imposed by the U(1)
$D$-term and  supersymmetry can be broken for arbitrary $N \geq 5$ as well.

The above conclusions 
can be trivially generalized for  arbitrary higher
dimensional couplings
\begin{equation}
W = \frac{\lambda\phi^n}{M_P^{n-1}}\Psi\bar{\Psi}\, .
\end{equation}
This will lead to the change  $\phi \rightarrow \phi^n/M_P^{n - 1}$ in the
 effective superpotential (\ref{efsuperw}). 

The question of whether 
the gauge contribution to the scalar soft masses
 dominate over the
gravity contribution
 is directly linked to the value of $\xi$. 
This is the subject on which we will
not speculate much here. Just note that,
{\it a priori}, the FI-term can be generated in the low-energy limit 
of some anomaly-free theory, 
after integrating out the heavy modes, 
provided that the light ones have Tr${\bf Q} \neq 0$. 
Obviously,  such a situation is impossible
within a four-dimensional field theory 
with unbroken supersymmetry, 
since the massive chiral superfields 
 appear in pairs with opposite charges
and do not contribute to 
Tr${\bf Q}$.
Nevertheless, the situation is different if 
 the theory changes dimensionality below the
integration scale, {\it e.g.}, 
if the four-dimensional  chiral theory
with Tr${\bf Q} \neq 0$ is obtained 
by compactifying some higher odd-dimensional one.
Since the original theory is anomaly free, the anomaly in the daughter
chiral theory must be cancelled by an effective Green-Schwarz term \cite{gs};
the quadratic divergence of the one-loop diagram that induces the $D$-term
must cancel out for  momenta larger than the inverse of the radius
of compactification $R^{-1}$ (the localization width in 
the extra dimension(s)).
The resulting FI-term then must be controlled by the radius 
$R^{-1}$ and can be small if $R$ is large.
In particular, 
in the M-theory picture \cite{stscale}, 
where the scales are more flexible,
the value of the FI-term may be smaller 
\footnote{We thank Michel Peskin for a
valuable discussion on this issue.}.

\subsection{Hybrid models of supersymmetry breaking}

If some of the SM fields are charged under the U(1),
the supersymmetry breaking
will be  transmitted by the U(1) and the gauge interactions.
Scalar masses (of the U(1)-charged fields) will arise from
  eq.~(\ref{dtermmass}) while  gaugino masses from 
eq.~(\ref{gauginos}).
We can see that
 there is a hierarchy of soft masses, 
$m_{Q_i}\gg m_\lambda$.
For  the scenario (a),
the scalar masses are two orders of magnitude larger
than the gaugino masses.
If only the first and second families
are charged under the U(1), this hierarchy can aminorate the
flavor and CP supersymmetric problem \cite{us}. 

If also the third  family  and 
the Higgs are charged, we  have in the scenario (a)
that either we take $m\simeq 100$ GeV 
and the gaugino masses are too small (see table~\ref{table}),
or we take $m={\cal O}(10$ TeV) and the weak scale  (that is related to the 
Higgs soft mass) is too large. 
Thus, a large degree of fine tuning is required in this case. 
In the scenario (b), this fine tuning  problem is aminorited
since the scalar masses of the fields charged under the U(1)
are only slightly larger
 than the gaugino masses (depending on the
value of $\varepsilon$ as can be seen in table~\ref{table}).
It is also possible to have
$m^2_H>m^2_Z$ in a natural way if
the SM Higgs arise as 
PGBs
from the breaking of some symmetry realized at high energies. 
As will see in the next section, 
this can be  easily implemented in
models with anomalous U(1).

\begin{table}
\begin{center}
\begin{tabular}{||c||c|c|c||}
\hline\hline
&$m_{Q_i}\  (q_i\not=0$)&$m_{\lambda}\, ,m_{Q_i}\  (q_i=0$)&
$m_{3/2}$\\
\hline\hline
 Scenario (a)     & $\sqrt{q_i}m$      
  &     $\sim10^{-2}m$
    &   $\sqrt{\varepsilon}m$      \\
\hline
Scenario (b)   &       $\sqrt{q_i}m$        
 &    $\sim10^{-2}\frac{m}{\sqrt{\varepsilon}}$ 
     &  $\sqrt{\varepsilon}m$       \\
\hline\hline
\end{tabular}
\end{center}
\caption[Table of masses]
{{
Soft masses for the scalar, gaugino and gravitino  generated in the 
scenarios (a) and (b) described in section~4.}}
\label{table}
\end{table}

\section{A  solution to the $\mu$-problem and 
  Higgs as pseudo-Goldstone bosons}

Let us suppose that the SM Higgs $H$ and $\bar H$ have the same
U(1)-charges as $\phi$ and $\bar\phi$ respectively.
In this case the superpotential (\ref{super})
can be extended to
\begin{equation}
  \label{superb}
W=m(\phi\bar\phi+\lambda H\bar H)\, .
\end{equation}
This will not alter the minimization procedure of section~2
and supersymmetry will be broken. {}From eq.~(\ref{superb})
one has 
a supersymmetric mass for the Higgs $\mu=\lambda m$ that results
to be of the right order of magnitude ({\it i.e.}, of the order
of the soft masses).
This mechanism for generating the $\mu$-term has different origin
from that in  ref.~\cite{mu}. There,
the  $\mu$-term is generated when supersymmetry
is broken.
Here, we start with  a   $\mu$-term before the breaking of
supersymmetry, and it is this term the  responsible for the generation
of the  soft masses.

Considering the case that $m$ is a  frozen parameter
(scenario (b)), we have 
that the Higgs  mass matrix (prior electroweak  breaking) is
given by
\begin{equation}
{\cal M}^2_{H\bar H}
=\left(\begin{array}{ll} \bar m^2_H&-B\mu\\  
-B\mu&\bar m^2_{\bar H}\\  
\end{array}\right)\, ,
\end{equation}
where up to ${\cal O}(m^2_{3/2})$,
\begin{eqnarray}
  \label{masses}
\bar m^2_H&\equiv&\mu^2+m^2_H=\lambda^2 m^2-m^2\, , \nonumber\\  
\bar m^2_{\bar H}&\equiv&\mu^2+m^2_{\bar H}=
\lambda^2 m^2+q_{\bar\phi}m^2\, ,\nonumber\\
B&=&0\, .
\label{bvalue}
\end{eqnarray}

Let us now consider the limit $\lambda=1$. In this  case one finds
that the scalar $H$ is a massless state.
This is in fact expected since in the limit $\lambda=1$, 
the superpotential
(\ref{superb}) and the $D$-term of the anomalous U(1)
posses a global 
  SU(3)$_L$ symmetry with $\bar\phi$, $\phi$ and 
the Higgs transforming as  triplets,
$(\phi\, ,H)\in {\bf 3}$ and 
$(\bar\phi\, ,\bar H)\in {\bf\bar 3}$.
When  $\phi$ and $\bar\phi$ get a VEV, the SU(3)$_L$ symmetry
is broken down to the  SU(2)$_L$ symmetry of the SM. 
There is a Goldstone boson
associated with this breaking that transforms as a doublet under
the SU(2)$_L$; this is the Higgs field $H$.
Of course, if the field $H$ is a  true Goldstone boson,
it will be massless to all orders and the 
electroweak symmetry will not be broken.
We will assume that the global SU(3)$_L$
symmetry is broken by the gauge
interactions and Yukawa couplings; it is just an accidental
symmetry of the superpotential (\ref{superb}).
In this case $H$ is just a pseudo-Goldstone boson (PGB)
 and will gain mass at  higher loop orders. 
Models with Higgs as  PGBs have been extensively considered
in the literature to solve the doublet-triplet splitting problem in
grand unify theories (GUTs) \cite{pgbmodels}.
In these GUTs the global SU(3)$_L$ is embedded in 
a  global SU(6).

The above scenario
 suggests that 
 the mass of $H$  can be smaller than $m$, and consequently
 $m_Z<m=\mu$. 
We want to analyze this point with more detail.
The minimization condition of the SM Higgs potential
is given by
\begin{equation}
  \label{ewsb}
\frac{m^2_Z}{2}=  
\frac{\bar m^2_{\bar H}-\bar m^2_{ H}\tan^2\beta}{\tan^2\beta-1}\, ,
\end{equation}
where
\begin{equation}
  \label{tanb}
 \sin 2\beta=\frac{2B\mu}{\bar m^2_H+ \bar m^2_{\bar H}}\, .
\end{equation}
For $q_{\bar\phi}\not=-1$, we have from eqs.~(\ref{bvalue}) that 
$B\mu\ll\bar m^2_{\bar H}$
and we are naturally   in the 
large $\tan\beta$ regime.
Notice that this is  different from models with gravity-mediated
supersymmetry breaking  where  one finds \cite{fine} that 
Higgs as PGBs usually leads to $\tan\beta$  close to one.
In the large $\tan\beta$ region,
eq.~(\ref{ewsb}) gives
\begin{equation}
  m^2_Z\simeq -2\bar m^2_H\, ,
\label{ewsbb}
\end{equation}
that relates the weak scale to the PGB mass.
At the scale $\sqrt{\xi}$, we already showed  that $\bar m_H$ is zero
at tree-level independently of the value of $\mu$. 
Nevertheless,
 it will receive radiative corrections. 
The largest corrections arises from
the scale evolution of $\bar m_H$ from
$\sqrt{\xi}$ to $m_Z$.
Up to corrections of ${\cal O}(m_{\lambda},A,m_{3/2})$ (that are small
in these models), we have 
from  section~3 that the soft mass of $H$, $m_H$, is not
modify; therefore   only the $\mu$-parameter will receive large 
corrections (a similar scenario is obtained in ref.~\cite{bim}
but in a different context). 
The $\mu$-parameter, being a supersymmetric
parameter, will be only renormalized by the wave-function renormalization
constants of $H$ and $\bar H$, {\it i.e.},
$\mu\rightarrow Z_HZ_{\bar H}\mu$.
Nevertheless, it is interesting to
notice that $Z_HZ_{\bar H}$ are slightly modified for values of
$m_t$ close to  the experimental values.
This is due to a partial cancellation between the (positive) gauge 
contribution and the (negative) top contribution.
Explicitly, one finds at the scale $m$:
\begin{equation}
  \bar m^2_H= \left[f^{\frac{3}{5b_1}}_1f^{\frac{3}{b_2}}_2
\sqrt{1-\frac{m_t^2}{m^2_{FP}}}-1\right]m^2
+{\cal O}(m^2_{3/2},m^2_\lambda)\, ,
\label{pgbmass}
\end{equation}
where
\begin{equation}
f_a=1+b_a\frac{\alpha_a(\sqrt{\xi})}{4\pi}\ln\frac{\xi}{m^2}\, , \ \ \ \
b_{1,2}=(33/5,1)\, ,
\end{equation}
$\alpha_a(\sqrt{\xi})$ is the gauge coupling at the scale $\sqrt{\xi}$;
 $m_t$ and 
$m_{FP}$ are respectively the running top mass and 
 the infrared fixed-point value for the   top mass
\footnote{It is defined as
$m^2_{FP}(t)=\frac{16\pi^2m^2_W}{3g^2}\frac{E(t)}{F(t)}\sin^2\beta$, where
the functions $E(t)$ and $F(t)$ are given in ref.~\cite{running}.}
at the scale $m$.
We have neglected the bottom contribution that could be important if 
$\tan\beta$ is very large.
We see that for  
\begin{equation}
  m^2_t=(1-f^{\frac{-6}{5b_1}}_1f^{\frac{-6}{b_2}}_2)m^2_{FP}\simeq 
(168\ {\rm GeV})^2\, \sin^2\beta\, ,
\label{value}
\end{equation}
the dominant contribution to $\bar m^2_H$ 
(the first term of 
eq.~(\ref{pgbmass})) cancels 
and 
the value of 
$\bar m^2_H$ remains of ${\cal O}(m^2_{3/2},m^2_\lambda)$.
This  suggests that for values of $m_t$ close to that of eq.~(\ref{value}),
we can have  $\mu=m\gg \bar m_H\sim m_Z\sim m_\lambda$.
Experimentally, we have \cite{topmass}
$m_t^{pole}=175.6\pm 5.5$ GeV that implies 
a running mass $m_t(m_t)=167.1\pm 5.2$ GeV; 
therefore the value of eq.~(\ref{value})
lays inside the experimental window.

In order to have
the right electroweak breaking without fine tuning,
we must require that
the value of
$|\bar m^2_H|$ 
coming from eq.~(\ref{pgbmass}) is
smaller then the weak scale
\footnote{ In fact, one could determine, using eq.~(\ref{ewsbb}),
the exact value of
$\mu$. For this purpose, however, one must include the gravitino and
gaugino contributions 
to the low-energy parameter $\bar m^2_H$;  one  should also consider
the one-loop effective potential 
instead of the tree-level one. 
Here we are only interested in obtaining a rough estimate
of the upper  bound on $m$.}, 
 {\it i.e.},
$|\bar m^2_H|\lappeq m^2_Z$. 
This  puts an upper bound on $\mu$.
In fig.~1 we plot this bound as a function of $m^{pole}_t$ and 
for $\tan\beta=10$. 
We have made the following approximations:
We have taken $\sqrt{\xi}=M_{GUT}\simeq 2\times 10^{16}$ GeV and
evaluated $\bar m^2_H$ at the scale $\approx 200$ GeV
where  $m_{FP}\simeq 196\times\sin\beta$ GeV.
We have  not included the 
effects on $\bar m^2_H$ arising from the
gaugino, trilinears or gravitino; 
 these effects are subdominant and very model dependent. 
{}From fig.~1 we see that large values of $\mu$ can be 
natural for certain values 
of the top mass.
Note that large values of $m$ also lead to heavy stop  and sbottom.
This is because 
the sum rule (\ref{sumrule}) implies
$m^2_{Q_3}+m^2_{U_3}=-m^2_H=m^2$ up to small radiative corrections
(we are assuming that the Yukawa coupling of the top  arises from
the tree-level term $h_tHQ_3U_3$).
Hence,
the only light superparticles in these models 
are the gauginos and gravitino.

If  $m$ 
in eq.~(\ref{superb}) is dynamical, the only difference from the case above
is that  $B\mu$ can be of the order of $m^2$ (see appendix) 
and  one can have smaller values 
of $\tan\beta$.
An upper  bound on $\mu$, 
similar to that of fig.~1, 
can be also obtained in this case. 
A complete analysis of the small $\tan\beta$ regime 
will be presented elsewhere.

\begin{figure}[htb]
\center{\mbox{\epsfig{file=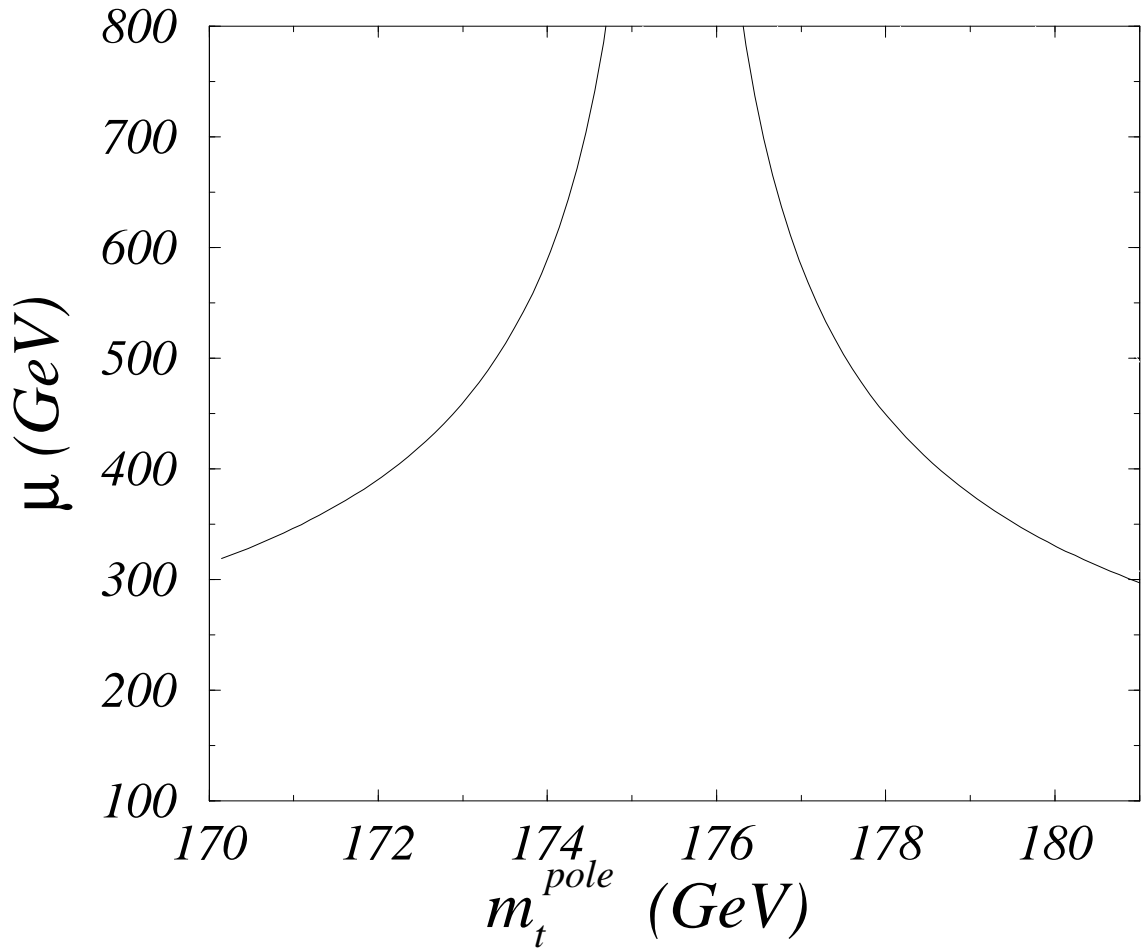,width=15.cm}}}
\caption{Upper bound on
 $\mu$ arising from  $|\bar m^2_H|\leq m^2_Z$,
as a function of $m^{pole}_t$ and for  $\tan\beta=10$.}
\label{fig1}
\end{figure}

\section{Conclusions}

The superparticle mass spectrum and its  future experimental test
 depends on the mechanism responsible for  supersymmetry breaking. 
It is then important to study  the different alternatives
that we know
to break supersymmetry and analyze its phenomenological consequences.

Here we have shown that models with anomalous U(1) allow for 
different possibilities to transmit the breaking of
supersymmetry to the SM fields. We have shown that:

\noindent$\bullet$ 
If the SM fields are neutral under the anomalous U(1),
then the model contain extra matter fields
that act as messenger of the breaking of supersymmetry. 
In this case, the pattern of soft masses are like that
in models of gauge-mediated supersymmetry breaking.
Gravity contribution can be important in these models,
depending on the value of the FI-term --see eq.~(\ref{nondeg}).

\noindent$\bullet$ 
When the SM fields are charged under 
the anomalous U(1), 
the soft masses present a 
 hierarchy of values (see table~1). 
{}For $\varepsilon\sim 10^{-3}$, we have:
\begin{equation}
{\rm Scenario (a)}:\ \ m_{Q_i}\gg m_{3/2}\gappeq m_{\lambda}\, ,
\label{hierarchya}
\end{equation}
\begin{equation}
{\rm Scenario (b)}:\ \ m_{Q_i}\gappeq m_{\lambda}>m_{3/2}\, ,
\label{hierarchyb}
\end{equation}
where the scenarios (a) and (b)  are defined in section~4.

\noindent$\bullet$ 
The soft masses of the fields charged under the anomalous U(1)
are (approximately)  RG invariant. This implies that,
even that supersymmetry is communicated to those fields at high energies, 
the soft masses  are not modified by the  
physics at the ultraviolet.
This provides a way to avoid
 the effect of ref.~\cite{hkr}.

\noindent$\bullet$ 
When the Higgs is charged under the anomalous U(1), 
the $\mu$-parameter is present in the theory with
 the right order of
magnitude.

\noindent$\bullet$ 
One of the Higgs of the SM can arise   as a  PGB.
In this case, its mass (before electroweak breaking) is smaller 
than the other soft masses since it is generated at the one-loop
order. As a consequence the weak scale is smaller than the $\mu$-parameter
and the supersymmetry breaking mass $m$
(see fig.~1).

\vspace{1cm}

It is a pleasure to thank 
Riccardo Barbieri, Savas Dimopoulos, Emilian Dudas, Michel Peskin, 
Riccardo Rattazzi and Carlos Wagner
for very useful discussions. 
A.P.  also wants to thank the CERN Theory Division for its hospitality.

\newpage

\noindent{\Large{\bf Appendix}}
\vspace{0.6cm}

\noindent
In this appendix we analyze the vacuum of a theory where the mass
$m$ is generated dynamically. 
Consider a SU(N) gauge group with $N_f$ 
 flavors, $\Phi_i$ and 
$\bar\Phi^j$ with  $i,j=1,...,N_f<N$. The gauge interactions 
becomes strong at some scale $\Lambda$, and
the quantum vacuum is described by the mesons 
$M^j_i\equiv\bar\Phi^j\Phi_i$ \cite{nonpert}. 
Let us assume that $M^j_i$ transforms under the anomalous U(1)
with charge $(1-q_{\bar\phi})$.
In this theory the superpotential is given by 
\begin{equation}
W=\frac{{\rm Tr} M}{M_P}\phi\bar\phi
+(N-N_f)\left(\frac{\Lambda^{3N-N_f}}
{{\rm Det}M}\right)^{\frac{1}{N-N_f}}\, ,
\label{superlast}
\end{equation}
where the second term is generated nonperturbatively \cite{ads,nsvz}.
The $F$-terms are given by
\begin{eqnarray}
F_\phi&=&\bar\phi\frac{{\rm Tr} M}{M_P}\, ,\label{fterma}\\
F_{\bar\phi}&=&\phi\frac{{\rm Tr} M}{M_P}\, ,\\
F_{M^j_i}&=&
\frac{\phi\bar\phi}{M_P}\delta^j_i-(M^{-1})^j_i
\left(\frac{\Lambda^{3N-N_f}}
{{\rm Det}M}\right)^{\frac{1}{N-N_f}}\, .
\label{ftermc}
\end{eqnarray}
Due to  the $D$-term  of the anomalous U(1) [eq.~(\ref{dtermcon})],
 $\phi$ is forced to get a VEV and therefore 
the $F$-terms of 
eqs.~(\ref{fterma})-(\ref{ftermc}) cannot all be zero, 
{\it i.e.},
supersymmetry is broken. 
The value of $m$ is given by
\begin{equation}
m=\frac{\langle{\rm Tr} M\rangle}{M_P}\, ,
\end{equation}
where $\langle M^j_i\rangle$, 
 for  $\Lambda\ll M_P$, is given by
\begin{equation}
\langle M^j_i\rangle
=\delta^j_i\left[\Lambda^{\frac{3N-N_f}{N}}
\left(\frac{\langle\phi\bar\phi\rangle}
{M_P}\right)^{\frac{N_f-N}{N}}\right]\, .
\end{equation}
Notice that 
since $\Lambda\ll \langle\phi\bar\phi\rangle/M_P\sim 
 \sqrt{\xi}$, one has 
$\langle M^j_i\rangle\ll \Lambda^2$ and 
the condensation takes place in the strong regime.
In this regime we do not know the dependence of the  K\"ahler
on $M$ and therefore we cannot calculate the exact value 
of the ratio $\langle\bar\phi\rangle/\langle\phi\rangle$ that 
depend on
the K\"ahler metric  $K_{MM}$:
\begin{equation}
\left(\frac{\langle\phi\rangle}
{\langle\bar\phi\rangle}+\frac{q_{\bar\phi}
\langle\bar\phi\rangle}{\langle\phi\rangle}\right)^{-1}
=-\frac{\langle K^{-1}_{MM}F_M\rangle}
{M_Pm^2(q_{\bar\phi}+1)}\, .
\label{last}
\end{equation}
Eq.~(\ref{last}) is derived from the a
minimization condition of the potential.
Assuming a normalized K\"ahler, we find (from other minimization
condition of the potential) 
\begin{equation}
\frac{\langle K^{-1}_{MM}F_M\rangle}{M_P}\simeq m^2\, ,
\label{fvev}
\end{equation}
and therefore 
\begin{equation}
\langle\bar\phi\rangle\simeq\langle\phi\rangle\, .
\label{equalvev}
\end{equation}
We then estimate that $\langle\bar\phi\rangle/\langle\phi\rangle$
lays between
$\sqrt{\varepsilon}$ and $1$, where the lower bound comes from gravity
contributions (see above eq.~(\ref{newvev})).
The upper and lower value of $\langle\bar\phi\rangle/\langle\phi\rangle$
 correspond to the 
scenario (a) and (b) respectively.

For the model of section~5, we just have to make the replacement
$\phi\bar\phi\rightarrow (\phi\bar\phi+ \lambda H\bar H)$ 
in eq.~(\ref{superlast}).
Unlike the case where $m$ was taken to be a frozen parameter,
eqs.~(\ref{bvalue}), we find that now $B\mu$ can be different from zero:
\begin{equation}
B\mu=\frac{\langle K^{-1}_{MM}F_M\rangle}{M_P}\, .
\end{equation} 
Due to the unknown strong effects on K\"ahler metric,
 we cannot 
determined $B\mu$ exactly; we can only estimate from eq.~(\ref{fvev}) 
 that its value is  close to  $m^2$.
For $\lambda=1$ we must have  a massless scalar (the PGB)
associated with the breaking of the global SU(3)$_L$ symmetry.
This implies that 
${\rm Det}{\cal M}^2_{H\bar H}=0$ and therefore 
$|B\mu|=\bar m_H\bar m_{\bar H}$.
We then see that 
the value of $B\mu$ can be  inferred (at tree-level)
from the PGB condition.
The  PGB is now the linear  combination 
\begin{equation}
\frac{1}{\sqrt{\bar m^2_H+\bar m^2_{\bar H}}}
\left(\bar m_{\bar H}H+ \bar m_H\bar H^\dagger\right)\, .
\end{equation}

\end{document}